\documentclass{aa}
\usepackage{times}

\begin{document}

\thesaurus{08(09.09.1 Crab nebula; 09.03.2; 09.19.2; 13.07.2)}

\title{COMPTEL $\gamma$-ray study of the Crab nebula}

\author{R.D.~van~der~Meulen\inst{2,5} 
  \and  H.~Bloemen\inst{2,5}
  \and  K.~Bennett\inst{4}
  \and  W.~Hermsen\inst{2}
  \and  L.~Kuiper\inst{2}
  \and  R.P.~Much\inst{4}
  \and  J.~Ryan\inst{3}
  \and  V.~Sch\"{o}nfelder\inst{1}
  \and  A.~Strong\inst{1}
         }

\institute{    
               Max-Planck-Institut f\"{u}r extraterrestrische
               Physik, P.O. Box 1603, 
               \mbox{D-85740 Garching}, Germany
\and
               SRON-Utrecht,
               Sorbonnelaan 2, \mbox{NL-3584 CA Utrecht},
               The~Netherlands               
\and
               Space Science Center, Univ. of New Hampshire,
               \mbox{Durham NH 03824}, U.S.A.
\and
               Astrophysics Division, ESTEC,
               \mbox{NL-2200 AG Noordwijk}, The~Netherlands
\and
               Leiden Observatory,
               P.O. Box 9513,
               \mbox{NL-2300 RA Leiden}, The~Netherlands
           }

\date{Received 19 December 1996 / Accepted 11 September 1997}

\offprints{R.~D.~van~der~Meulen (R.vanderMeulen@sron.ruu.nl)}


\maketitle
\begin{sloppypar}
\begin{abstract}
We report on a study of the $\gamma$-ray continuum emission from the
Crab supernova nebula 
and on a search for nuclear de-excitation $\gamma$-ray lines. 
Crab is the brightest continuum source in the 1--10 MeV $\gamma$-ray sky, 
and its continuum radiation is most likely of synchrotron origin. 
It is a likely source of cosmic rays through shock acceleration 
and thus a potential candidate for $\gamma$-ray line emission from nuclear interactions. 
Five years of COMPTEL observations enable a fine spectral binning 
to investigate the behaviour of the 0.75--30 MeV emission in detail 
and to search for nuclear de-excitation lines on top of the continuum. 
The nebular spectrum shows a break 
at the edge of the COMPTEL energy range 
and connects well to the EGRET spectrum, 
probably reflecting electron energy losses in the synchrotron emission scenario.
Such a smooth continuum model alone may not be sufficient to explain the observations.
A weak bump in the spectrum at 1--2 MeV may be present.
No significant evidence for distinct line emission is seen,
but the presence of a blend of line features or another synchrotron component
cannot be excluded.\\


\end{abstract}

\section{Introduction}

Strong shocks in Galactic supernova remnants (SNRs) are widely believed
to be sources of cosmic rays (CRs) below $10^{15}$ eV.
This has only been substantiated, however,
through detection of accelerated energetic electrons 
by their synchrotron emission. 
Direct evidence for the acceleration of the nuclear component 
is still to be found.
Searches for signatures of the nuclear component in the TeV range 
(e.g. Atoyan \& Aharonian \cite{atoyan}; Allen et al. \cite{allen}), 
for instance, have only provided upper limits.
The existence of accelerated protons and nuclei can also be deduced 
from the presence of $\gamma$-ray lines which these particles produce 
when interacting with the ambient medium.

Nuclear interaction line emission has so far been reported from solar flares
(e.g. Chupp \cite{chupp}; Murphy et al. \cite{murphy}; 
Ryan et al. \cite{ryan}) 
and in only one occasion from an extra-solar source, 
the Orion Complex (Bloemen et al. \cite{bloemwij}, \cite{bloemby}). 
Evidence is seen now in emission from the inner Galaxy as well
(Bloemen et al. \cite{bloemin}). 
These detections were obtained with the Compton Telescope (COMPTEL) 
aboard the Compton Gamma Ray Observatory (CGRO). 
COMPTEL operates between 0.75 and 30 MeV, 
which is the energy regime where signs of nuclear interactions can be expected.
The observed 3--7 MeV emission from Orion was tentatively attributed to 
$^{12}$C$^*$ (4.44 MeV) and $^{16}$O$^*$ (6.13 MeV) de-excitation lines. 
Solar flare spectra show several additional lines 
from heavier nuclei ($^{20}$Ne, $^{24}$Mg, $^{28}$Si),
mainly in the 1--2 MeV regime.
Here we report on a search for such $\gamma$-ray lines from the Crab nebula, 
using COMPTEL data obtained during the first five years of the mission.

The Crab nebula is an extensively studied 943 year old SNR.
The pulsar wind from the central, isolated pulsar 
supplies high-energy particles that synchrotron radiate 
(Shklovsky \cite{shklov}; Dombrovsky \cite{dombrov}; Oort \& Walraven \cite{oort}) 
while gyrating in the magnetic field of the nebula (Kennel \& Coroniti \cite{kennel}).
However, in situ acceleration has to occur as well.
The radio-to-$\gamma$-ray spectrum consists of a mix of different power-law components,
separated by breaks (e.g. Zombeck \cite{zom}; De Jager et al. \cite{dejager}), 
which seem to indicate three electron populations. 
The continuum emission up to $\sim\!1.8\times10^{13}$Hz
is associated with the optically visible nebula.
The break may reflect the energy losses for electrons as old as the SNR;
the average field of the optical nebula can then be calculated to be
$\sim\!3\times10^{-4}$ G (e.g. Trimble \cite{trimble}; Marsden et al. \cite{marsden}). 
Consequently, the electrons responsible for the torus of X-ray emission, 
the spectrum of which breaks somewhere between 60 and 150 keV (Bartlett \cite{bartlett2}),
should have a lifetime of $\sim\!1$ month.
The inner edge of the nebula lies at 8" (0.08 pc), 
so it is clear that these electrons have been accelerated 
in or near the nebula.

De Jager et al. (\cite{dejager}) discuss evidence for a third spectral break 
at $\sim\!26$ MeV, from early observations by COMPTEL and EGRET (also aboard CGRO),
which is addressed below.
This would indicate another population of energetic electrons, 
with lifetimes of $\sim\!1$ day. 
We present further evidence for this spectral break 
using more COMPTEL and EGRET observations.

\begin{table}[t]
\label{tab:obs}
\caption[]{COMPTEL observations used in the present Crab analysis. 
           The observation numbers are given in standard CGRO notation.}
\scriptsize
 \begin{flushleft}
 \begin{tabular}{cccccc}
 \hline
Obs &  T{$_{\rm start}$} & T{$_{\rm end}$} & Obs &  
T{$_{\rm start}$} & T{$_{\rm end}$} \\
\# & dd-mm-yy & dd-mm-yy & \# & dd-mm-yy & dd-mm-yy \\
\hline
0$^a$ & 28-04-91 & 07-05-91 & 321.5 & 15-02-94 & 17-02-94  \\
1.0 & 16-05-91 & 30-05-91 & 337 & 09-08-94 & 29-08-94  \\
31 & 11-06-92 & 25-06-92 & 412 & 28-02-95 & 07-03-95 \\
36.0 & 11-08-92 & 12-08-92 & 413 & 07-03-95 & 21-03-95 \\
36.5 & 12-08-92 & 20-08-92 & 419.1 & 04-04-95 & 11-04-95 \\
39.0 & 01-09-92 & 17-09-92 & 419.5 & 09-05-95 & 23-05-95 \\
213 & 23-03-93 & 29-03-93 & 420 & 23-05-95 & 06-06-95 \\
221 & 13-05-93 & 24-05-93 & 426 & 08-08-95 & 22-08-95 \\
310 & 01-12-93 & 13-12-93 & 502 & 17-10-95 & 31-10-95 \\
321.1 & 08-02-94 & 15-02-94 &&& \\ 
 \hline
 \multicolumn{6}{l}{$^a$ Data obtained during the verification phase}
\end{tabular}
\end{flushleft}
\normalsize
\end{table}

\section{Instrumentation, observations \& data analysis}

\subsection{Instrument}
\label{sec:instr}
COMPTEL is designed to detect $\gamma$-ray photons with energies 
between about $0.75$ and $30$ MeV with an energy resolution of 5--10\% FWHM. 
It has a field-of-view of $\sim\!1$ steradian 
and an angular resolution of typically 1\degr--3\degr, 
which enables the instrument to monitor several objects simultaneously. 
The location accuracy is about $0.5^\circ$ for a strong source. 
A detailed description of the detection principle and instrument 
is given by Sch\"{o}nfelder et al.\ (\cite{schoen}). 
Ideally, incoming $\gamma$-ray photons are first Compton scattered 
in an upper detector layer and then completely absorbed
in a lower detector layer. 
The measured energy deposits and locations in these layers determine 
the scatter direction, scatter angle, and total energy of each photon. 
For selected energy bands, the telescope events are binned 
in a 3-dimensional data space, 
which consists of two scatter direction coordinates 
and a Compton scatter angle coordinate. 
Here we use a binning of $1^\circ\/\times\/1^\circ\/\times\/ 1^\circ$.
The source response function of the instrument in this data cube 
has a cone-like shape and depends on the actual source spectrum. 
In the present work we have adopted an $E^{-2}$ power-law input spectrum, 
but our findings are not sensitive to this specific choice.

\begin{table}[t]
\label{tab:flux}
\caption[]{Crab nebula (as plotted in Fig.~1) 
           and total (Fig.~3d) fluxes.}
\small
 \begin{flushleft}
\begin{tabular}{crrcrr}
\hline
 E & Flux$_{\rm unp}$ & Error$_{\rm unp}$ && Flux$_{\rm tot}$ & 
Error$_{\rm tot}$ \\
MeV & \multicolumn{2}{c}{$10^{-5} \gamma$ cm$^{-2}$ s$^{-1}$} && 
\multicolumn{2}{c}{$10^{-5} \gamma$ cm$^{-2}$ s$^{-1}$}   \\[0.5ex]
 \hline 
0.78--0.96 & 34.07 & 1.57 && 42.75 & 1.00 \\
0.96--1.16 & 31.11 & 1.30 && 36.68 & 0.82 \\
1.16--1.38 & 21.54 & 1.09 && 27.80 & 0.69 \\
1.38--1.62 & 18.68 & 1.04 && 23.78 & 0.65 \\
1.62--1.88 & 13.03 & 0.82 && 17.02 & 0.52 \\
1.88--2.16 & 8.90 & 0.71 && 11.62 & 0.45 \\
2.16--2.48 & 7.84 & 0.70 && 10.25 & 0.44 \\
2.48--2.84 & 8.66 & 0.57 && 9.79 & 0.36 \\
2.84--3.22 & 6.14 & 0.52 && 7.28 & 0.33 \\
3.22--3.62 & 4.41 & 0.46 && 6.13 & 0.29 \\
3.62--4.08 & 4.28 & 0.41 && 5.80 & 0.26 \\
4.08--4.56 & 4.44 & 0.36 && 4.77 & 0.22 \\
4.56--5.08 & 2.94 & 0.33 && 3.96 & 0.21 \\
5.08--5.66 & 3.42 & 0.32 && 3.69 & 0.20 \\
5.66--6.26 & 2.30 & 0.28 && 2.91 & 0.18 \\
6.26--6.94 & 2.15 & 0.28 && 2.64 & 0.18 \\
6.94--7.64 & 1.60 & 0.25 && 2.11 & 0.16 \\
7.64--8.42 & 1.17 & 0.20 && 1.80 & 0.13 \\
8.42--9.26 & 1.36 & 0.17 && 1.75 & 0.11 \\
9.26--10.16 & 1.24 & 0.15 && 1.47 & 0.10 \\
10.00--11.20 & 1.05 & 0.15 && 1.57 & 0.10 \\
11.20--12.48 & 1.12 & 0.14 && 1.37 & 0.09 \\
12.48--13.92 & 1.12 & 0.14 && 1.24 & 0.09 \\
13.92--15.52 & 0.62 & 0.13 && 0.92 & 0.09 \\
15.52--17.28 & 0.83 & 0.14 && 0.96 & 0.09 \\
17.28--19.28 & 0.50 & 0.14 && 0.68 & 0.09 \\
19.28--21.60 & 0.31 & 0.14 && 0.69 & 0.10 \\
21.60--24.08 & 0.23 & 0.15 && 0.29 & 0.10 \\
24.08--26.88 & 0.16 & 0.18 && 0.26 & 0.11 \\
26.88--30.00 & 0.13 & 0.27 && 0.09 & 0.18 \\
\hline
 \end{tabular}
\end{flushleft}
\normalsize
\end{table}

\begin{figure*}[tb]
\label{fig:unp}
\vspace{9.5cm}
\caption[]{COMPTEL spectrum of the Crab nebula, 
           together with GRIS low-energy data (Bartlett et al. \cite{bartlett1}) 
           and EGRET high-energy data (Fierro \cite{fierro}).  
           Up to 10 MeV, the bin width of the COMPTEL data points is equal to
           twice the FWHM of the instrumental energy resolution. 
           Between 10 and 30 MeV, broader energy bins were chosen 
           (Sect.~\ref{sec:instr}). 
           The two COMPTEL upper limits at the highest energies
           (Table~2) are not shown.}
\includegraphics{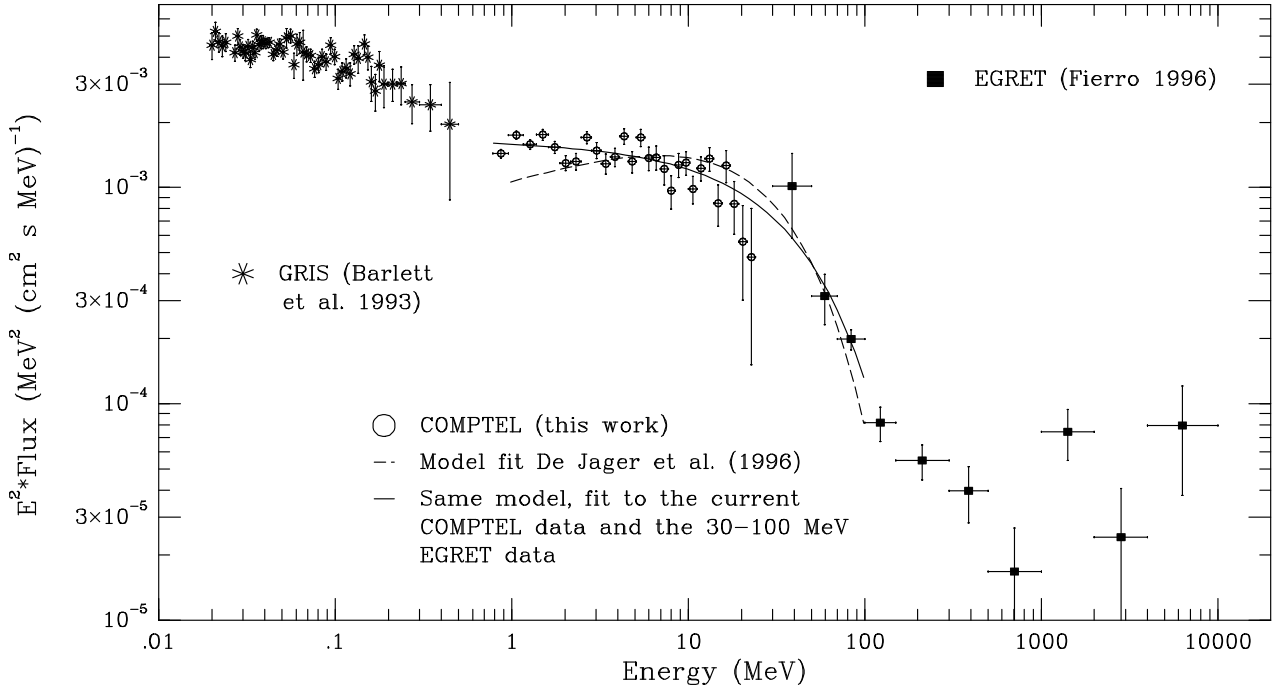}
\end{figure*}

We applied a maximum-likelihood method (de Boer et al. \cite{deboer}) 
to obtain flux estimates 
and to construct spectra with narrow energy bins
equal to twice the FWHM of the COMPTEL energy resolution, 
given by $${{\rm{FWHM}}(E)= 0.0236 \times (14.61E + 2.53E^2)^{1/2}}$$
\noindent with $E$ in MeV (Sch\"{o}nfelder et al. \cite{schoen}). 
Our method of an\-alysis does not allow a significantly finer energy binning
at this stage.
At high energies ($> 10$ MeV) the limited number of events 
requires even larger bins. 
In this study we have a total of 30 statistically independent bins.

On average, more than $95$\% of the number of events 
tagged by the instrument as Compton-scattered $\gamma$-ray 
should be attributed to the background (instrumental and isotropic). 
The likelihood analysis requires, 
for each energy interval separately, 
a careful estimate of this background 
for which a variety of methods has been studied.
The results shown in this paper are based on 
a method of background determination 
in which a filter technique is applied to the data space. 
This method is described by Bloemen et al.\ (\cite{bloemher}), 
although we applied here an improved algorithm 
involving an iterative process of background estimation. 
The filter eliminates to first order any source signature present.
In each iteration the background is further corrected 
for the smeared-out source signature.

In this work we use point spread functions (PSFs) from analytical modeling
based on single-detector calibrations. PSFs from Monte Carlo simulations
of the instrument are preferred, but not available yet for the narrow
energy bins used here. A globally somewhat softer spectrum is
expected with the simulated response (about 0.1 in spectral index).

\subsection{Observations}
Crab was within 30\degr of the instrument pointing 
during several observations in the first 5 years of CGRO operations 
(Table~\ref{tab:obs}). Observation 2.5 was excluded because the instrumental settings were optimized for solar flare observations.

\subsection{Phase selection}

In addition to the Crab nebula, the Crab pulsar is a strong (pulsed)
$\gamma$-ray source as well, 
which fully dominates the high-energy ($\ga 100$ MeV) $\gamma$-ray emission
(e.g. Kanbach et al. \cite{kanbach}; Nolan et al. \cite{nolan}; 
Fierro \cite{fierro}).
In the COMPTEL energy range, about 20\% of the total Crab emission
is estimated to be due to the pulsar (cf Fig. 6 in Much et al. \cite{muchben1}).
COMPTEL cannot resolve the Crab pulsar from the surrounding nebula spatially,
but it is possible to disentangle them by pulsar
phase selection, assuming that there
is no pulsar emission in the ``off''-phase at $\gamma$-ray energies.
We determined the unpulsed emission by selecting
the events in the off-pulse phase 0.525-0.915, as defined by
Nolan et al. (\cite{nolan}).
Our analysis thus gives 39\% of the flux, which is then normalized 
to the full period.
The observations listed in Table~\ref{tab:obs} were combined and folded by our
pulsar analysis software. 

\section{Results}

\begin{figure}[t]
\label{fig:simunp}
\vspace{14cm}
\caption[]{a) 0.75--9.7 MeV COMPTEL spectrum of the Crab nebula (solid lines). 
         The bin sizes are twice the FWHM of the instrumental energy resolution
         and shifted by 1 FWHM with respect to the points in (b), also shown (dotted).
         b) 0.78--10.2 MeV nebula spectrum as in Fig.~1;
         the triangles indicate positions of some candidate nuclear interaction lines :
         $^{24}$Mg (1.37 MeV),
         $^{20}$Ne (1.63 MeV),
         $^{28}$Si (1.78 MeV), 
         {$^{12}$C} (4.44 MeV), and 
         {$^{16}$O} (6.13 MeV) (Ramaty et al. \cite{ramaty})
         c,d,e) Simulations of E$^{-2}$ spectra at a level of 
         $1.38 \times 10^{-3}$ MeV cm$^{-2}$ s$^{-1}$.}
\includegraphics{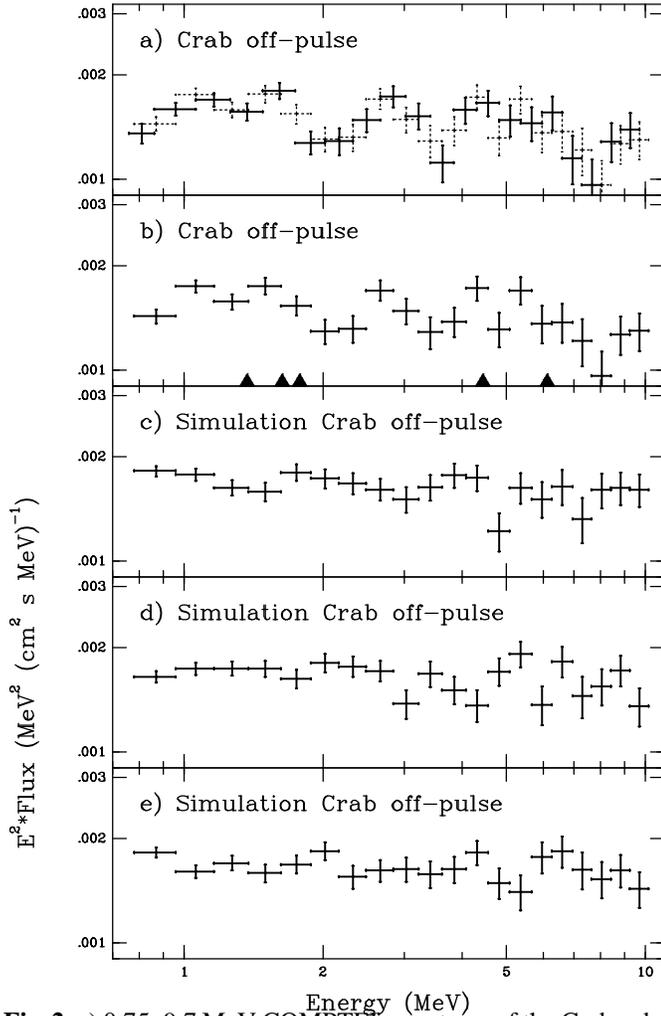}  
\end{figure}

\begin{figure}[tb]
\label{fig:simtot}
\vspace{19cm}
\caption[]{a) 0.75--9.7 MeV COMPTEL spectrum of the total Crab emission (solid lines). 
         The bin sizes are twice the FWHM of the instrumental energy resolution
         and shifted by 1 FWHM with respect to the points in (d), also shown (dotted).
         b,c,d) 0.78-10.2 MeV total Crab spectrum
         using observations from respectively the first three years, 
         the fourth and fifth, and the first five years of operations.
         e,f,g) Simulations of E$^{-2}$ spectra
         at a level of $1.75 \times 10^{-3}$ MeV cm$^{-2}$ s$^{-1}$.}
\includegraphics{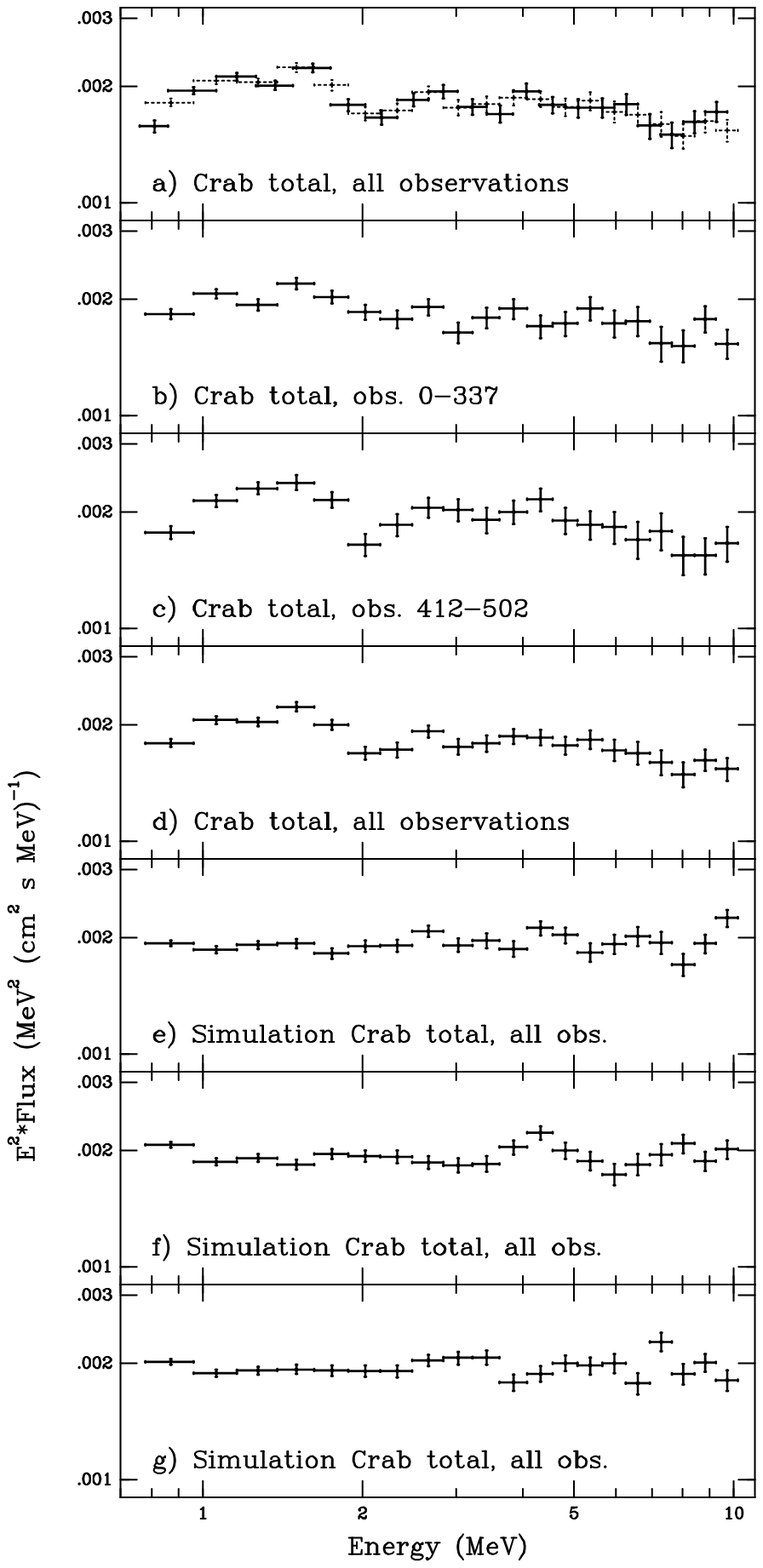}  
\end{figure}

Figure~1 shows a power-per-decade spectrum of the off-pulse fluxes
(Tab.~2). 
Our data points are in good agreement
with the COMPTEL spectrum presented by Much et al. (\cite{muchben2}), 
which was based on observations obtained during the first two years 
of the mission and contains 7 data points only.
Statistical error bars are shown. 
The absolute calibration uncertainty is conservatively estimated to be 30\%. 
The total number of photons in this spectrum is $\sim\/68000$.
Measurements from GRIS (Bartlett et al. \cite{bartlett1}) 
and EGRET (Fierro \cite{fierro}, April 1991 -- Aug. 1994) 
are added to place the COMPTEL points in a broader perspective.

De Jager et al. (\cite{dejager}) have modelled the emission of the Crab nebula.
They used an inverse-Compton (IC) component and a
synchrotron component with an exponential cut-off 
to fit COMPTEL (Much et al. \cite{muchben2}) and EGRET data 
(both April 1991 --  May 1993):
$${dN/dE = K_{\rm{s}}(E/3.5{\rm{MeV}})^{-{\Gamma}_{\rm{s}}}{\rm{exp}}(-E/E_0)}$$
$${+ K_{\rm{IC}}(E/1000 {\rm{MeV}})^{-\Gamma_{\rm{IC}}}}.$$

\noindent
Their best-fit result is shown as a dashed line in Fig.~1
(limited to the 1--100 MeV range). 
Figure~1 shows that the spectral break represented in the model 
begins in the COMPTEL domain.
We have fit our data and the first 3 EGRET points
to the first term of the equation. 
We ignore the IC component of the model, 
which is valid because the IC influence on the used data points is negligible.
For the fit parameters we find: 
the energy cut-off $E_0 = 41 \pm 3$ MeV, 
$K_{\rm{s}} = (1.29 \pm 0.04)\times 10^{-4}$ cm$^{-2}$ s$^{-1}$ MeV$^{-1}$, 
and $\Gamma_{\rm{s}} = 2.02 \pm 0.03$, 
with a reduced $\chi^2$ ranging from 1.9 to 2.2 
for 28 degrees of freedom (dof). The latter
$\chi^2$ value was obtained by fitting the model to a spectrum with bins
shifted by 1 FWHM of the energy resolution.  Figure~2a shows, in solid lines,
this shifted spectrum from 0.75--10 MeV. For comparison, the unshifted
spectrum, also shown in Fig.~2b, is represented by dotted lines.
The $1\sigma$ errors on the fit parameters have been calculated 
by increasing the $\chi^2_{\rm{min}}$ with 3.5 (Lampton et al. \cite{lampton}).
The result can be seen in Fig.~1 (solid line).
The cut-off energy is somewhat sensitive to the EGRET data points. 
Leaving all EGRET points out gives fit parameters of 
$E_0 = 20^{+3}_{-2}$ MeV, 
$K_{\rm{s}} = (1.47 \pm 0.04)\times 10^{-4}$ cm$^{-2}$ s$^{-1}$ MeV$^{-1}$, and $\Gamma_{\rm{s}} = 1.90 \pm 0.03$, 
with a reduced fit $\chi^2$ ranging from 2.0 to 2.3, for 25 dof. 
For comparison, De Jager et al. (\cite{dejager}), 
who included the COMPTEL calibration uncertainty in their fit 
but excluded the $<1$ MeV COMPTEL point, 
found: 
$E_0 = 26^{+26}_{-9}$ MeV, 
$K_{\rm{s}} = 1.25\times 10^{-4}$ cm$^{-2}$ s$^{-1}$ MeV$^{-1}$, and $\Gamma_{\rm{s}} = 1.74 \pm 0.42$.

The fit is not perfect, as indicated by the $\chi^2$ value obtained above,
but no obvious systematic trends can be seen. 
The deviations from the model fit are
below the 3$\sigma$ level in the individual energy bins. 
In order to assess the systematic errors of our method 
we analyzed three Monte Carlo simulations of the Crab up to 10 MeV (Fig.~2cde),
assuming an E$^{-2}$ spectrum 
with an integrated 1--10 MeV flux of $1.38 \times 10^{-3}$ cm$^{-2}$ s$^{-1}$,
as derived from the observations.
It appears that with this input flux
(corrected for the phase selection, i.e. 
$0.39 \times 1.38 \times 10^{-3}$ cm$^{-2}$ s$^{-1}$) 
our method generates randomly placed features 
similar to those in the observed spectrum. 
Therefore, no significance can be given to the observed features.
Best fits of the simulations to power-law spectra 
$${dN/dE = K(E/3.5{\rm{MeV}})^{-{\Gamma}}}$$
\noindent give $\chi^2_{\rm{red}}$ values of 0.9, 1.4, and 0.9 respectively,
for 18 dof. 
For comparison, the best power-law fit to the observed data up to 10 MeV 
has an index of $2.02^{+0.03}_{-0.02}$, 
$K=$($1.12\pm 0.03$)$ \times 10^{-4}$ cm$^{-2}$ s$^{-1}$, 
and a $\chi^2_{\rm{red}}$ ranging from 2.6 to 3.0, 
depending on the bin positions. 
The $1\sigma$ errors on the fit parameters have been calculated 
by increasing the $\chi^2_{\rm{min}}$ with 2.3 (Lampton et al. \cite{lampton}).

Since a mixture of line features (including broad-line components)
may well be present, 
it cannot be excluded that we overestimated the continuum contribution.  
On the other hand, a decrease of the fitted continuum level
would not exhibit clear evidence for known candidate lines 
(triangles in Fig.~2b)
without introducing other features as well (e.g. near 2.8 MeV).
Gravitationally redshifted $\gamma$-ray lines caused by 
ions falling onto the pulsar 
have not been included in the candidate lines, 
because it is unknown how deep in the atmosphere the particles would interact,
and thus how large the redshift would be.

Much et al. (\cite{muchben1}) reported that the Crab unpulsed emission 
might be time variable.
Spectral changes (e.g. the position of the break energy) 
may occur on short time scales, 
possibly affecting the global appearance of our spectrum.
Time variability of the Crab emission will be the topic of future study.

In order to obtain the cleanest data, we selected only 39\% 
(the unpulsed fraction) of the available nebula events. 
However, as the emission from the nebula dominates 
the total emission from the Crab, 
we have used the set of all available (pulsar contaminated) events 
for our analysis as well. 
Figure~3d shows the total Crab spectrum using all observations; Fig.~3a
shows the same spectrum (dotted lines) and the spectrum with shifted bins
(solid lines). 
The integrated  1--10 MeV flux 
of $1.75\times 10^{-3}$ cm$^{-2}$ s$^{-1}$ 
has been used as input parameter for simulations 
of an E$^{-2}$ spectrum (Fig.~3efg).  
The best power-law fits to the simulations have 
$\chi^2_{\rm{red}}$ values of 1.5, 2.4 and 1.7 (18 dof). 
Clearly, this is not perfect; 
the deviations may result from the fact that
we have used the actual observations as the basis for our simulations,
so that the smoothed Crab signal may influence our findings.
Anyway, the best power-law fit to the observed data up to 10 MeV 
[$\Gamma = 2.02\pm 0.01$, 
 $K= $($1.41\pm 0.02$)$ \times 10^{-4}$ cm$^{-2}$ s$^{-1}$], 
has a $\chi^2_{\rm{red}}$ ranging from 5.0 to 7.4 
(depending on the bin positions), 
which is a clear indication that the spectrum is more complicated.
The main difference between observation and simulations 
is in the low energy range; at 1--2 MeV there is a hint for a broad feature.
Some evidence for this feature could in fact already be seen in the
unpulsed spectrum of Fig.~1, but it seems more pronounced in the total
Crab spectrum.
After splitting the data in two time intervals 
(Fig.~3bc, observations 0-337 and 412-502 respectively), 
some evidence for this feature is seen in both spectra.
We cannot exclude, however, that it is associated with the pulsar,
if real.
The BATSE (35--1700 keV) Cycle 1--3 total Crab spectrum (Ling \cite{ling}), 
although higher than the GRIS and COMPTEL points (Much et al. \cite{muchhar}),
shows a deviation from a simple power-law spectrum, 
which can be interpreted as a rise near 1 MeV as well.  
If real, the feature could well be due to a blend of lines, or it could be
the synchrotron signature of a different population of electrons.
Morphological comparisons of OSSE, BATSE and COMPTEL Crab spectra 
will be the topic of future study.

\section{Conclusions}

COMPTEL observations from the first five years of the mission 
were combined to search, in a fine binned spectrum,
for nuclear de-excitation $\gamma$-ray lines from the Crab supernova nebula. 
No significant evidence for line emission is found.
A reasonable fit to the off-pulse fluxes can be obtained 
with the synchrotron component of the model from 
De Jager et al.\ (\cite{dejager}),
giving a break energy of about 25--40 MeV.
An indication for a bump at 1--2 MeV cannot readily be
explained by known
systematic and statistical errors.
It is possible that line contributions or an extra continuum component
have resulted in an overestimation of the single continuum component.

\acknowledgements  
The COMPTEL project is supported by the German government through DARA grant 
50 QV 90968, by NASA under contract NAS5-26645 and by NWO. 
R.D. van der Meulen is supported by the Netherlands Foundation for Research in
Astronomy (NFRA) with financial aid from the Netherlands Organization for 
Scientific Research (NWO).

\end{sloppypar}
\end{document}